\documentclass[prd,twocolumn,nofootinbib,reprint]{revtex4-1}
\usepackage{amsmath,amssymb,amsthm,bm,braket,ascmac,bbm}
\usepackage{graphicx}
\usepackage{color} 

\theoremstyle{definition}

\begin{document}
\title{Footprints of New Strong Dynamics via Anomaly and the 750 GeV Diphoton}
\author{Yuichiro Nakai$^1$, Ryosuke Sato$^{2,3}$ and Kohsaku Tobioka$^{2,3,4}$}
\affiliation{\vspace{2mm} $^1$Department of Physics, Harvard University, Cambridge, Massachusetts 02138, USA \\
$^2$Department of Particle Physics and Astrophysics, Weizmann Institute of Science, Rehovot 7610001, Israel \\
$^3$Institute of Particle and Nuclear Studies, High Energy Accelerator Research Organization (KEK),
Tsukuba 305-0801, Japan \\
$^4$Raymond and Beverly Sackler School of Physics and Astronomy,
Tel-Aviv University, Tel-Aviv 69978, Israel}


\begin{abstract}
\vspace{1mm}

The chiral anomaly provides a smoking-gun evidence of a new confining gauge theory.
Motivated by a reported event excess in a diphoton invariant mass distribution at the LHC,
we discuss a scenario that a pseudo-Nambu-Goldstone (pNG) boson of a new QCD-like theory
is produced by gluon fusion and decays into a pair of the standard model gauge bosons.
Despite the strong dynamics, the production cross section and the decay widths are determined by anomaly matching condition.
The excess can be explained by the pNG boson with mass of around 750 GeV.
The model also predicts exotic hadrons such as
a color octet scalar and baryons.
Some of them are within the reach of the LHC experiment.

\end{abstract}
\maketitle

\section{Introduction}\label{sec:intro}

A new confining gauge theory is ubiquitous in physics beyond the standard model (SM).
Most models to solve the hierarchy problem, such as technicolor
\cite{Hill:2002ap}, (the holographic picture of) Randall-Sundrum scenarios
\cite{Randall:1999ee}
and a class of little Higgs models
\cite{Katz:2003sn}, involve new strong dynamics.
Even in supersymmetry
\cite{Martin:1997ns}, the breaking scale is often assumed to be given by dimensional transmutation of a new gauge theory
\cite{Witten:1981nf}.
In addition, the dark matter can be like a new pion originating from strong dynamics
\cite{Hochberg:2014kqa}.
Furthermore, string theory seems to prefer non-minimal models with extra gauge groups.
Therefore, from the viewpoint of such prevailing nature of strong dynamics,
even if we do not intend to solve the hierarchy problem,
it is well motivated to
pursue a possibility of a new asymptotically free gauge theory to be explored at the LHC.\footnote{
As a scenario with the similar motivation, vector-like confinement was studied in Ref.~\cite{Kilic:2009mi}.  For a different confinement scale, there is Hidden Valley scenario \cite{Strassler:2006im} (see also, Refs~\cite{Strassler:2006qa,Nakai:2015swg}).
}

Recently, the ATLAS and CMS collaborations have reported an excess in diphoton invariant mass distribution
around $m_{\gamma\gamma} \simeq 750~{\rm GeV}$ \cite{cern20151216}.
If this peak comes from a new scalar boson $\phi$ with mass of around 750 GeV,
the reported event number can be fitted with a relatively large production cross section times branching ratio
$\sigma(p p \to \phi) {\rm Br}(\phi \to \gamma\gamma) = 1.4$--$18~{\rm fb} $
(see the discussion in Section III).
Then, the scalar $\phi$ must be efficiently produced, which is achieved by gluon fusion.
The relevant interaction terms of $\phi$ are parametrized as
\begin{equation}
\begin{split}
{\mathcal L}_{\rm eff} &=
\frac{\alpha}{4\pi} \frac{k_\gamma}{\Lambda_\gamma} \phi \, F_{\mu\nu} \tilde F^{\mu\nu} 
+ \frac{\alpha_s}{4\pi} \frac{k_g}{\Lambda_g} \phi \, G^{a}_{\mu\nu} \tilde G^{a\mu\nu},
\label{Leff}
\end{split}
\end{equation}
where $F$ denotes the field strength of the photon, $\tilde F^{\mu\nu}\equiv \frac{1}{2}\epsilon^{\mu\nu\rho\sigma}F_{\rho\sigma}$, and $G$ indicates the gluon field strength.
$k_\gamma$ and  $k_g$ are dimensionless constants and $\Lambda_\gamma$, $\Lambda_g$ are mass parameters.
From the effective interactions of Eq.~\eqref{Leff}, the widths of $\phi$ decays into $gg$ and $\gamma\gamma$ are calculated as~\cite{Raffelt:2006cw}
\begin{equation}
\begin{split}
\!\Gamma(\phi \to gg) = \frac{\alpha_s^2}{8\pi^3} \frac{k_g^2 m_\phi^3}{\Lambda_g^2},\ \Gamma(\phi \to \gamma\gamma) = \frac{\alpha^2}{64\pi^3} \frac{k_\gamma^2 m_\phi^3}{\Lambda_\gamma^2},
\end{split}
\end{equation}
where $m_\phi$ is the mass of $\phi$.
 With a natural assumption $k_\gamma / \Lambda_\gamma \sim k_g / \Lambda_g$, the scalar boson $\phi$ dominantly decays into two gluons,
and the total decay width is approximately given by $\Gamma_{\phi} \simeq \Gamma(\phi \to gg)$.
We can also take the branching ratio of diphoton as
${\rm Br}(\phi\to\gamma\gamma) \simeq \Gamma(\phi \to \gamma\gamma) / \Gamma(\phi \to gg)$.
By using the narrow width approximation \cite{Georgi:1977gs,Cahn:1983ip},
the production cross section times branching ratio is then estimated as
\begin{align}
&\sigma(pp \to \phi + X ){\rm Br}(\phi \to \gamma\gamma)\simeq \frac{\pi^2}{8m_\phi s}
\Gamma(\phi \to \gamma\gamma) \nonumber\\[1ex]
&\quad \times \int_0^1 dx_1 \int_0^1 dx_2 \left[ 
\delta( x_1x_2 - {m_\phi^2}/{s}) g(x_1) g(x_2) \right],
\label{analyticcross}
\end{align}
where $s$ is the square of center-of-mass energy and $g(x_\alpha)$ ($\alpha = 1,2$) is the parton distribution function (PDF) of the gluon.
We evaluate it numerically by using the MSTW PDF \cite{Martin:2009iq}.
For $m_\phi = 750~{\rm GeV}$, the quantity of Eq.~\eqref{analyticcross} is given by
\begin{equation}
\begin{split}
&\sigma(pp \to \phi + X ){\rm Br}(\phi \to \gamma\gamma) \\[1ex]
&\quad\simeq \left( \frac{\Lambda_\gamma/k_\gamma}{100~{\rm GeV}} \right)^{-2}
\times 
\left\{
\begin{array}{ll}
1.8~{\rm fb} & (8~{\rm TeV}) \\[1.5ex]
8.6~{\rm fb} & (13~{\rm TeV})
\end{array}
\right.,
\label{numcross}
\end{split}
\end{equation}
where we have presented both cases of $\sqrt s = 8$ and $13 \, \rm TeV$.

\begin{figure}[!t]
  \begin{center}
          \includegraphics[clip, width=7cm]{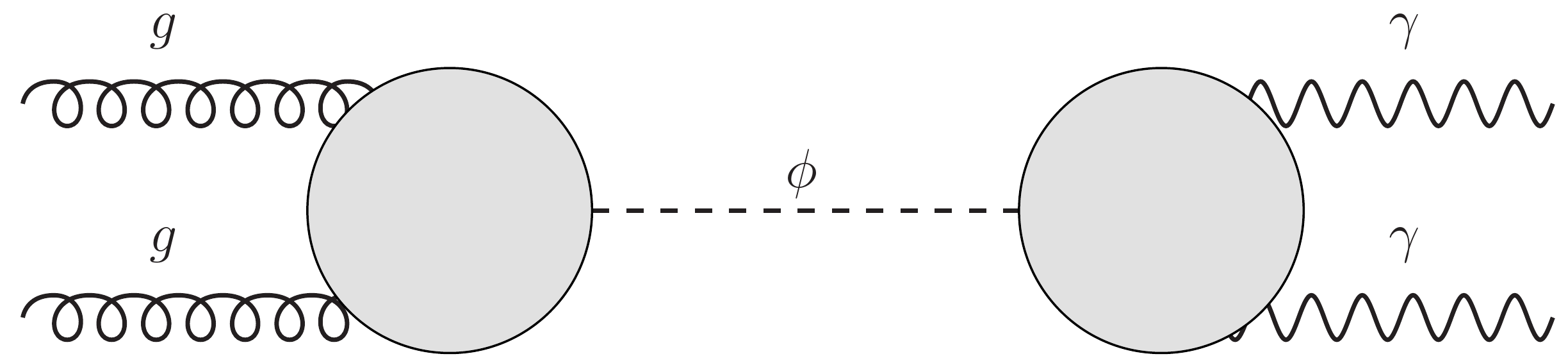}
    \caption{A new scalar $\phi$ is produced by gluon fusion and decays into two photons. In our model, the effective interactions are from the chiral anomaly.}
    \label{fig:diphoton}
\vspace{-0.5cm}
  \end{center}
\end{figure}

If the effective interactions of Eq.~\eqref{Leff} come from loops of some heavy particles as  in Fig.~\ref{fig:diphoton},
$\Lambda_\gamma, \Lambda_g$ $(\gtrsim m_\phi)$ are given by typical mass scales of the heavy particles.
For fitting the reported event number, Eq.~\eqref{numcross} indicates that
the combination of the parameters $\Lambda_\gamma / k_\gamma$ must be around $100~{\rm GeV}$.
Since the scalar mass is $m_\phi \simeq 750 \, \rm GeV$,
it is suggested that the dimensionless constant $k_\gamma$ is much larger than one.
That is, we need some strong dynamics to explain the reported diphoton excess.
Alternatively, we can consider a scenario in which $\phi$ is produced by quark anti-quark fusion via e.g., $(1/\Lambda) q_L q_R^c H \phi $.
However, basically the discussion is parallel and we need a large effective coupling $k_\gamma/\Lambda_\gamma \sim (100~{\rm GeV})^{-1}$.

In this paper, motivated by the reported event excess,
we explore a possibility of a new QCD-like theory to appear at the TeV scale.
The effective theory after confinement contains pseudo-Nambu-Goldstone (pNG) bosons
of an approximate chiral symmetry.
We discuss LHC phenomenology of these pNG bosons.
The lightest pNG boson is produced by gluon fusion and decays into a pair of the SM gauge bosons as in Fig.~\ref{fig:diphoton},
which explains the reported excess.
Importantly, despite the strong dynamics, the production cross section and the decay widths are determined
by the 't~Hooft's anomaly matching condition
like $\pi^0 \to \gamma \gamma$ in the ordinary QCD.
The excess can be explained by the pNG boson with mass of around 750 GeV.
Unlike most cases, the lightest pNG boson produced by gluons  has a photon-enriched signal as the second dominant decay. 
As discussed in the technicolor models \cite{Dimopoulos:1979sp, Dimopoulos:1980yf},
the model also predicts exotic hadrons such as color octet and triplet scalars and
baryons some of which are within the reach of the LHC experiment.

The rest of the paper is organized as follows.
In section $2$, we present a model with a new confining gauge interaction.
The masses of the pNG bosons and their effective interactions with the SM gauge bosons after confinement are analyzed.
In section $3$, we discuss collider phenomenology of the lightest pNG boson 
and explain the reported event excess.
In section $4$, phenomenology of exotic hadrons such as color octet and triplet scalars and baryons is described.
We make some concluding remarks in section $5$. 

\section{A new QCD}

Let us consider an asymptotically free $SU(N)$ gauge theory
with new Weyl fermions, $\psi, \bar \psi$ (color triplets) and $\chi, \bar \chi$ (color singlets), which are (anti-)fundamental under the $SU(N)$.
Their charge assignments are summarized in Table~\ref{tab:fermions}.
The new fermions have vector-like masses,
\begin{equation}
\begin{split}
{\mathcal L} \supset -M_\psi \psi \bar \psi - M_\chi \chi \bar \chi \, ,
\end{split}
\end{equation}
where $M_\psi$ and $M_\chi$ are mass scales of around $100 \, \rm GeV$.
When we neglect the $SU(3)_C$ and $U(1)_Y$ gauge couplings
and the mass scales $M_\psi$, $M_\chi$,
the theory has a global $SU(4)_{L} \times SU(4)_{R} \times U(1)_B \times U(1)_A$ symmetry
where $U(1)_B$ is the baryon number symmetry and $U(1)_A$ is anomalous
under the $SU(N)$ gauge interaction.

\begin{table}[!t]
\centering
\begin{tabular}{|c|c|c|c|c|}
\hline
         & $SU(3)_C$ & $SU(2)_L$ & $U(1)_Y$ & $SU(N)$ \\\hline
$\psi$      & $\bf 3$         & $\bf 1$          & $-1/3$   & $\bf N$ \\
$\chi$      & $\bf 1$         & $\bf 1$          & 1        & $\bf N$ \\
$\bar \psi$ & $\bf \bar 3$  & $\bf 1$          & $1/3$    & $\bf \bar N$ \\
$\bar \chi$ & $\bf 1$          & $\bf 1$          & $-1$     & $\bf \bar N$ \\\hline
\end{tabular}
\caption{
The charge assignments of the new fermions.
}\label{tab:fermions}
\vspace{-1mm}
\end{table}

The considered $SU(N)$ gauge theory is asymptotic free and confines at low energies.
As in the case of the ordinary QCD, the new fermions condense,
\begin{equation}
\begin{split}
\langle \psi \bar \psi \rangle  \sim \frac{4 \pi }{\sqrt{N}} f_S^3 \,  {\bf 1}, \qquad
\langle \chi \bar \chi \rangle \sim \frac{4 \pi }{\sqrt{N}} f_S^3 \, ,
\end{split}
\end{equation}
where $f_S$ is the decay constant and we have used naive dimensional analysis (NDA) for counting the factors of $4 \pi$ and $N$
(see e.g., Ref.~\cite{Luty:1997fk,Ponton:2012bi}).
Then, the approximate $SU(4)_L\times SU(4)_R$ global symmetry is broken down to the diagonal subgroup $SU(4)_V$,
in which the SM $SU(3)_C$ and $U(1)_Y$ gauge groups are embedded.\footnote{
If we took $M_\psi=M_\chi=0$ and the hypercharges of the new fermions as zero,
the Lagrangian would be the same as that of the Kim's composite axion model
\cite{Kim:1984pt}. 
The authors of Refs.~\cite{Cacciapaglia:2015nga,Cai:2015bss} considered similar models where matter fermions have different representations of the SM gauge groups.
In their models, the decays of the pNG bosons lead to $WW, ZZ$ rich signals.}
Associated with the chiral symmetry breaking,
there are 15 pNG bosons as light degrees of freedom and they behave under the $SU(3)_C$ as
$\bf 15 \to 8 + 3 + \bar 3 + 1$.
In the following discussion, we denote the $SU(3)_C$ octet, triplet and singlet pNG bosons
as $\phi_8$, $\phi_3$ and $\phi$ respectively.

\subsection{The masses of the pNG bosons}

We now estimate the masses of the pNG bosons by using chiral perturbation theory
(for a review, see \cite{Gasser:1984gg}).
The dependence on the mass parameters $M_\psi$, $M_\chi$ is determined by group theory,
while the squared masses of the pNG bosons are also proportional to an undetermined mass scale of order one 
in the unit of the dynamical scale $\Lambda_S$.
Then, we estimate the squared mass of the singlet pNG boson
by scaling up the formula for the QCD pion mass $m_\pi$.
The result is given by
\begin{equation}
\begin{split}
m_\phi^2
&\simeq \sqrt{\frac{3}{N}} \frac{M_\psi/4 + 3M_\chi/4}{(m_u + m_d)/2} \frac{f_S}{f_\pi} \, m_\pi^2 \\[1ex]
&\simeq (750~{\rm GeV})^2 \times \sqrt{\frac{3}{N}}  \left( \frac{M_\psi/4 + 3M_\chi/4}{25~{\rm GeV}} \right) \\[1ex]
&\qquad\qquad\qquad\times \left( \frac{f_S}{400~{\rm GeV}} \right)\left( \frac{(m_u+m_d)/2}{3.5~{\rm MeV}} \right) ,
\end{split}
\end{equation}
where we have used $\Lambda_S \sim 4 \pi f_S / \sqrt{N}$ from NDA and the QCD pion decay constant is $f_\pi \simeq 93~{\rm MeV}$.

Once we fix the mass of the singlet pNG boson $m_\phi$, the masses of the colored pNG bosons
are determined by chiral perturbation theory and radiative corrections from the ordinary QCD.
The squared masses of the triplet and octet pNG bosons are estimated as 
\begin{equation}
\begin{split}
&m_{\phi_3}^2 = \frac{M_\psi/2 + M_\chi/2}{M_\psi/4 + 3M_\chi/4} m_\phi^2 + \delta m_3^2 \, , \\[1ex]
&m_{\phi_8}^2 = \frac{M_\psi}{M_\psi/4 + 3M_\chi/4} m_\phi^2 + \delta m_8^2 \, .
\end{split} \label{eq:pNGBmass1}
\end{equation}
Radiative corrections from QCD  
\cite{Farhi:1980xs,Dobrescu:1996jp} are 
\begin{equation}
\begin{split} \label{eq:pNGBmass2}
\delta m_3^2 &~\simeq~  \frac{4}{3} \Delta
~\simeq~ (650~{\rm GeV})^2 \times \frac{3}{N} \left(\frac{f_S}{400~{\rm GeV}}\right)^2 \, , \\[1.5ex]
\delta m_8^2 &~\simeq~ 3 \Delta
~\simeq~ (970~{\rm GeV})^2 \times \frac{3}{N} \left(\frac{f_S}{400~{\rm GeV}}\right)^2 \, , \\[1.5ex]
\Delta &~\equiv~ \frac{\alpha_s(\Lambda_S)}{\alpha (\Lambda_{\rm QCD})} \frac{\Lambda_S^2}{\Lambda_{\rm QCD}^2} (m_{\pi^\pm}^2 - m_{\pi^0}^2)\, , 
\end{split}
\end{equation}
where the gauge couplings are evaluated at their respective dynamical scales,
$m_{\pi^0}$ and $m_{\pi^\pm}$ are the neutral and charged pion masses and 
we have used $\Lambda_S / \Lambda_{QCD} \sim \sqrt{3/N}f_S / f_\pi $ from NDA.
Because of the QCD radiative correction, the colored pNG bosons obtain large masses compared to the singlet pNG boson.
Furthermore, when the color triplet fermion $\psi$ is heavier than the singlet $\chi$, $M_\psi \gtrsim M_\chi$, 
the colored pNG bosons obtain large masses from the explicit chiral symmetry breaking.
Hereafter, we focus on a situation where the lightest new hadron is the singlet pNG boson $\phi$.

\subsection{The effective interactions}

To discuss LHC phenomenology of the pNG bosons,
we here present their effective interactions with the SM gauge bosons. 
It is important that the coefficients of these interactions are determined
by the 't~Hooft's anomaly matching condition
like $\pi^0 \to \gamma \gamma$ in the ordinary QCD (see e.g. Ref.~\cite{Weinberg:1996kr}).
We now write the pNG bosons collectively as
$\pi_s \equiv \phi_8^a T^a_8 + \phi_3^i T^i_3 + (\phi_3^i  T^i_3 )^\dagger + \phi  T_1$,
where $a=1,\dots,8$ and $i = 1, \dots, 3$ are the indices of the adjoint and fundamental representations 
of the $SU(3)_C$ gauge group.
The generators of the $SU(4)$ are given by
\begin{align}
&T_8^a = \frac{1}{2} \left(\begin{array}{cc}
\lambda^a & 0_{1\times 3} \\
0_{3\times 1} & 0
\end{array}\right),~ T_1 = \frac{1}{2\sqrt{6}} \left(\begin{array}{cc}
1_{3\times 3} & 0_{1\times 3} \\
0_{3\times 1} & -3 
\end{array}\right) \, , \nonumber\\
&(T_3^k)_{IJ} = \delta_{4I} \delta_{Jk} /\sqrt{2} \, .
\end{align}
Here, $\lambda^a_{ij}~(a=1,\dots,8)$ are the Gell-Mann matrices of the $SU(3)_C$
and $I, J = 1, \dots , 4$.
Then, the chiral anomaly leads to the effective Lagrangian of the pNG bosons $\pi_s$
interacting with the SM gauge bosons without precise knowledge of particles inside the loops,
known as Wess-Zumino-Witten anomaly
\cite{Wess:1971yu, Witten:1983tw, Witten:1983tx}.
The result is given by
\begin{equation}
\begin{split}
{\mathcal L}_{\rm eff} = -\frac{N}{8\pi^2 f_S} {\rm tr}[\pi_s {\mathcal F}_{\mu\nu} \tilde {\mathcal F}^{\mu\nu}] \, , \label{eq:phiFFtilde}
\end{split}
\end{equation}
where the field strength $\mathcal{F}$ is defined as
${\mathcal F}_{\mu\nu} \equiv g_s G^a_{\mu\nu} T^a_8 + g_Y B_{\mu\nu} Q_Y$.
Here $G$ and $B$ are the field strengths of the $SU(3)_C$ and $U(1)_Y$ gauge groups,
$g_Y$ is the $U(1)_Y$ gauge coupling and 
$Q_Y \equiv {\rm diag} (-1/3,-1/3,-1/3,1)$.
With the effective interaction Eq.~\eqref{eq:phiFFtilde},
we can discuss LHC phenomenology of the pNG bosons as we will see next.

\section{The lightest pNG boson}

We now consider LHC phenomenology of the singlet pNG boson 
and show that  the reported excess can be explained by this boson $\phi$ with mass of $750 \, \rm GeV$.
We can extract the interactions of $\phi$ with the SM gauge bosons from Eq.~(\ref{eq:phiFFtilde}) as
\begin{align}
&{\mathcal L}_{\phi \mathcal{F}  \mathcal{\tilde F}}
=
- \frac{N}{\sqrt{6}} \frac{\phi}{f_S} \frac{g_s^2}{32\pi^2} G_{\mu\nu} \tilde G^{\mu\nu} 
+ \frac{16N}{3\sqrt{6}} \frac{\phi}{f_S} \frac{e^2}{32\pi^2} F_{\mu\nu} \tilde F^{\mu\nu}
\nonumber\\[1ex]
&~  - \frac{32N}{3\sqrt{6}} \frac{\phi}{f_S} \frac{e^2 t_W}{32\pi^2} F_{\mu\nu} \tilde Z^{\mu\nu} 
+ \frac{16N}{3\sqrt{6}} \frac{\phi}{f_S} \frac{e^2 t_W^2}{32\pi^2} Z_{\mu\nu} \tilde Z^{\mu\nu} ,
\label{lightest}
\end{align}
where $Z_{\mu\nu}$ is the field strength of the $Z$ boson, $e$ is the electric charge and we have defined $t_W \equiv \tan \theta_W$
for simplicity.
Then, we can compare these effective interactions with the Lagrangian of Eq.~\eqref{Leff},
which leads to
$k_\gamma/\Lambda_\gamma = 8N / (3 \sqrt{6} f_S)$ and
$k_g/\Lambda_g = -N/(2 \sqrt{6} f_S)$.
Note that we can take $(k_\gamma / \Lambda_\gamma)^{-1}$ to be around 100 GeV by choosing $f_S$.
This is consistent with $m_\phi = 750 \, \rm GeV$
for appropriate values of $M_\psi$, $M_\chi$ by virtue of new strong dynamics.
\begin{figure}[t!]
\centering
\includegraphics[width=1\linewidth]{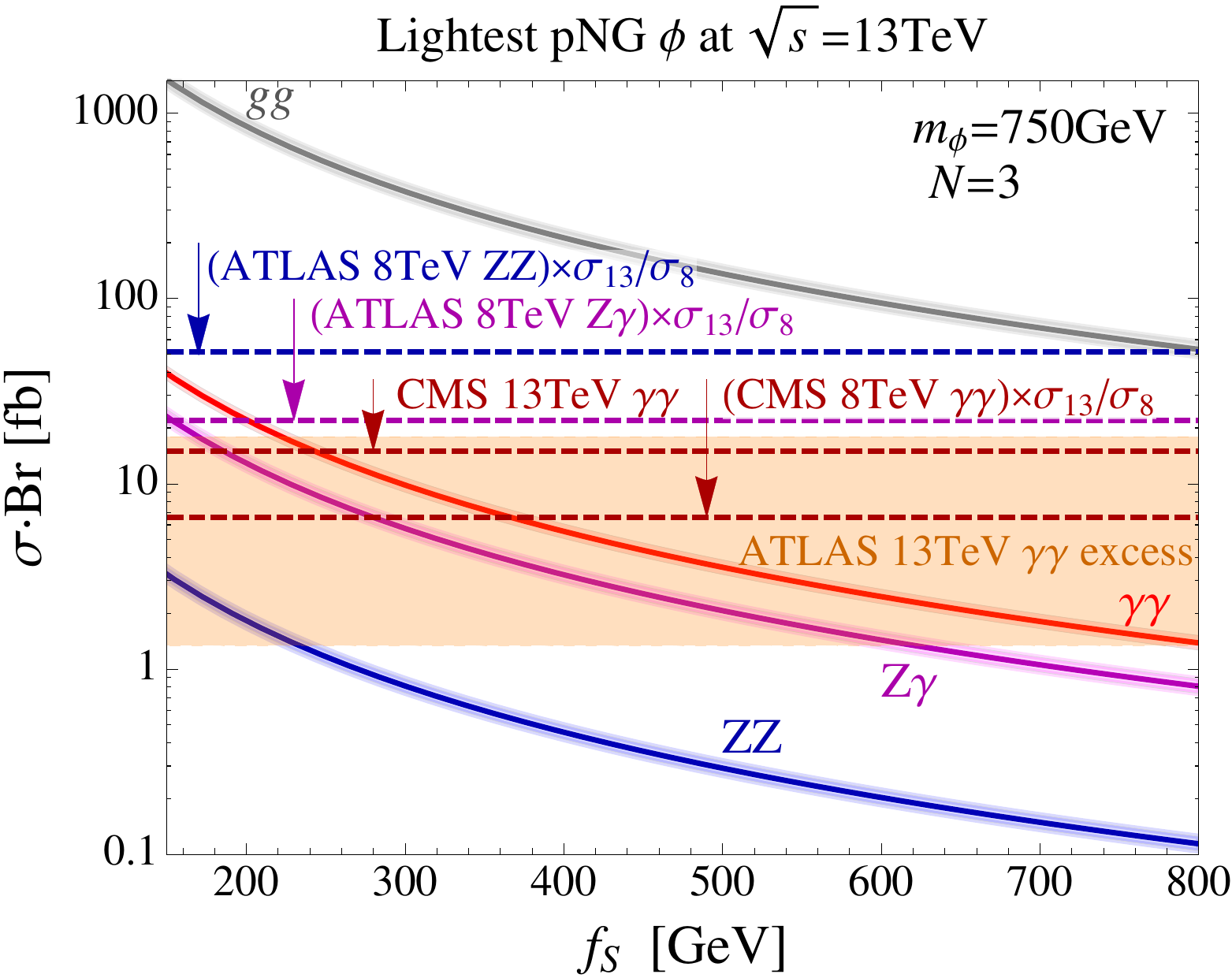}
\caption{The lightest pNG boson signal expected at the LHC  run of $\sqrt{s}$=13~TeV. The lines are $\sigma\cdot$Br of $\phi\to gg, \gamma\gamma, Z\gamma, ZZ$ from a gluon fusion production with $m_\phi=750$~GeV.  Here $SU(3)$ new strong dynamics is adopted. PDF uncertainty is given by a band along with each line. Upper bounds (95\%~C.L.) from resonance searches at $\sqrt{s}$=8 TeV in $\gamma\gamma$ \cite{Khachatryan:2015qba}, $ZZ$ \cite{Aad:2015kna} and $Z(\to ll)\gamma$ \cite{Aad:2014fha} final states are rescaled by a ratio of cross section, $\sigma^{\rm ggF}_{\rm 13TeV}/\sigma^{\rm ggF}_{\rm 8TeV}$, and plotted as dashed lines. For $\sqrt{s}$=13~TeV results, the orange shaded region is consistent with an excess of $\gamma\gamma$ resonance reported by ATLAS \cite{ATLAS-CONF-2015-081} while CMS places an upper bound in the similar analysis (red dashed line)\cite{CMS-PAS-EXO-15-004}.
\label{fig:phi13TeV}
}
\end{figure}

Given the effective interactions, the lightest pNG boson is efficiently produced with gluon fusion and decays into a pair of the SM gauge bosons
at the LHC.
The decay widths of $\phi$ are calculated as
\begin{equation}
\begin{split}
\Gamma(\phi \to gg)   &= \frac{N^2 \alpha_s^2 m_\phi^3}{192 \pi^3 f_S^2}, \ 
\Gamma(\phi \to \gamma\gamma) = \frac{N^2 \alpha^2 m_\phi^3}{54 \pi^3 f_S^2}, \\[1ex]
\Gamma(\phi \to \gamma Z) &= \frac{N^2 \alpha^2 t_W^2 m_\phi^3}{27 \pi^3 f_S^2} \left(1-\frac{m_Z^2}{m_\phi^2}\right)^3 , \\[1ex]
\Gamma(\phi \to ZZ)   &= \frac{N^2 \alpha^2 t_W^4 m_\phi^3}{54 \pi^3 f_S^2} \left(1-\frac{4m_Z^2}{m_\phi^2}\right)^{3/2},
\end{split}
\end{equation}
where $m_Z$ is the $Z$ boson mass.
Although these widths depend on $f_S$, their ratios are independent.
When we fix the mass of $\phi$ as $m_\phi = 750~{\rm GeV}$, 
the ratios are numerically given by
\begin{equation}
\begin{split}
&\Gamma(gg): \Gamma(\gamma\gamma): \Gamma(\gamma Z): \Gamma(ZZ) \\[1ex]
\simeq
& \, 0.965 : 0.021 : 0.012 : 0.002 \, .
\end{split}
\end{equation}
We can see that the decay to two gluons is dominant and the decay to diphoton is next.
By using the MSTW PDF \cite{Martin:2009iq}, we obtain
the production cross section times branching ratio for the decays to two gluons and photons
(up to NLO uncertainty which is beyond the scope of the present paper),
\begin{equation}
\begin{split}
&\sigma(pp \to \phi + X ){\rm Br}(\phi \to gg) \\[1ex]
&\simeq \left(\frac{N}{3}\right)^2 \left( \frac{f_S}{400~{\rm GeV}} \right)^{-2}
\times 
\left\{
\begin{array}{ll}
56~{\rm fb} & (8~{\rm TeV}) \\[1ex]
220~{\rm fb} & (13~{\rm TeV})
\end{array}
\right., \\[1.5ex]
&\sigma(pp \to \phi + X ){\rm Br}(\phi \to \gamma\gamma) \\[1ex]
&\simeq \left(\frac{N}{3}\right)^2 \left( \frac{f_S}{400~{\rm GeV}} \right)^{-2}
\times 
\left\{
\begin{array}{ll}
1.2~{\rm fb} & (8~{\rm TeV}) \\[1ex]
5.8~{\rm fb} & (13~{\rm TeV})
\end{array}
\right..
\end{split}
\end{equation}
We have shown the cases of $\sqrt s =$ 8 and 13 TeV for each decay.

In Fig.~\ref{fig:phi13TeV}, we show the lightest pNG boson signal expected at the LHC  run of $\sqrt{s}$=13~TeV. The lines are $\sigma\cdot$Br of $\phi\to gg, \gamma\gamma, Z\gamma, ZZ$ from a gluon fusion production with $m_\phi=750$~GeV.  The PDF uncertainty is given by a band along with each line. The CMS diphoton resonance search with  integrated luminosity $\cal L=$ 2.6~fb$^{-1}$ places an upper bound,  $\sigma\cdot{\rm Br}_{\gamma\gamma}\lesssim 15$~fb \cite{CMS-PAS-EXO-15-004}. Based on 
an excess in the corresponding search of ATLAS with $\cal L=$ 3.2 fb$^{-1}$ \cite{ATLAS-CONF-2015-081}, we estimate the allowed region in the following way. We take the two bins nearest to 750~GeV (Fig.~1 in Ref.~\cite{ATLAS-CONF-2015-081}) which have 23 events and 10.5 expected background. Using a log-likelihood ratio constructed from Poisson probability distribution function, we find the allowed number of signals $S$ is from 4.3 to 23 events at 95\%~C.L. Systematic uncertainties are assumed to be subdominant since the statistical error is large. When it is rescaled to $\sigma\cdot$Br, we vary the acceptance times efficiency, ${\cal A}\cdot\epsilon=$ 40--100\%, and therefore we obtain 
	\begin{align}
	1.4 ~{\rm fb}\lesssim \sigma\cdot{\rm Br}_{\gamma\gamma}=\frac{S}{ {\cal L} ({\cal A}\cdot\epsilon)}\lesssim 18~{\rm fb}
	\text{ (95\% C.L.).}
	\end{align}
These results of ATLAS and CMS are consistent, and the excess can be explained by $f_S(N/3)^{-1}=250$--800 GeV.
Other upper bounds are obtained by relevant resonance searches at $\sqrt{s}$=8 TeV in $\gamma\gamma$ \cite{Khachatryan:2015qba}, $ZZ$ \cite{Aad:2015kna}, and $Z(\to ll)\gamma$ \cite{Aad:2014fha}, which are rescaled by a ratio of $\sigma^{\rm ggF}_{\rm 13TeV}/\sigma^{\rm ggF}_{\rm 8TeV}\approx 4.7$ in Fig.~\ref{fig:phi13TeV}. 
For  the $Z(\to ll)\gamma$ bound  \cite{Aad:2014fha}, the signal efficiency within the fiducial volume is assumed to be 80\%.   

If the diphoton signal will be established, the next evidence is expected in $Z\gamma$ resonance. Based on the current sensitivity~\cite{Aad:2014fha}, the prospect at $\sqrt{s}$=14~TeV estimated by rescaling the gluon parton luminosity is  that the interesting parameter space of $f_S(N/3)^{-1}\lesssim800$~GeV can be proven with 300~fb$^{-1}$ at 95\%C.L.

\section{Exotic hadrons}

In addition to the singlet pNG boson $\phi$,
there are the color octet ($\phi_8$) and triplet ($\phi_3$) pNG bosons.
The model also has baryonic states whose masses are around a TeV.
Some of them are within the reach of the LHC experiment.
In this section, we discuss the phenomenology of the color octet scalar and also comment on the other bound states.

\subsection{The octet pNG boson}

We here consider the color octet pNG boson $\phi_8$.
As in the case with the lightest pNG boson,
the octet $\phi_8$ has the interaction terms with the SM gauge bosons from the chiral anomaly,
which can be read from Eq.~(\ref{eq:phiFFtilde}) as
\begin{align}
{\mathcal L}_{\phi_8 \mathcal{F} \mathcal{ \tilde F}}
=&
- \frac{N g_s^2}{32\pi^2 f_S} \phi_8^a G_{\mu\nu}^b \tilde G^{c\mu\nu} d_{abc} 
- \frac{N g_s e}{24\pi^2 f_S} \phi_8^a G_{\mu\nu}^a \tilde F^{\mu\nu}  \nonumber\\[1ex]
&+ \frac{N g_s e \, t_W}{24\pi^2 f_S} \phi_8^a G_{\mu\nu}^a \tilde Z^{\mu\nu}  \, .
\end{align}
Here, we have used $\{\lambda^b, \lambda^c \} = (4/3)\delta_{bc} + 2 d_{abc} \lambda^a$ and ${\rm tr}[\lambda^a \lambda^b] = 2\delta^{ab}$.
The decay widths of $\phi_8$ into two SM gauge bosons are then calculated as
\begin{align}
\Gamma(\phi_8 \to gg) &
= \frac{5}{4} \frac{m_{\phi_8}^3}{m_\phi^3} \Gamma(\phi \to gg) ,\nonumber\\
\Gamma(\phi_8 \to g\gamma) &
= \frac{2}{3}\frac{\alpha}{\alpha_s} \frac{m_{\phi_8}^3}{m_\phi^3} \Gamma(\phi \to gg) ,\\
\Gamma(\phi_8 \to gZ) &
= \frac{2}{3}\frac{\alpha t_W^2}{\alpha_s} \frac{m_{\phi_8}^3}{m_\phi^3} \left(1-\frac{m_Z^2}{m_{\phi_8}^2}\right)^3 \Gamma(\phi \to gg) \, ,\nonumber
\end{align}
where $\sum_{b,c} d_{abc} d_{bcd} = (5/3)\delta_{ad}$ has been used.
When we take $m_\phi = 750~{\rm GeV}$ and $m_{\phi_8} = 1300~{\rm GeV}$,
the ratios of the decay modes are numerically given by 
$\Gamma(gg): \Gamma(g\gamma): \Gamma(g Z) \simeq \, 0.943 : 0.044 : 0.013$\ .
We can see that the decay to two gluons is dominant and the decay to a gluon and a photon is next.

In addition to the decay modes to the SM gauge bosons, the octet scalar 
can decay into $\phi$ because $\phi_8$ is heavier than the singlet $\phi$.
While $\phi_8 \to \phi g$ is prohibited by angular momentum conservation \cite{sakurai},
the decay mode $\phi_8 \to \phi gg$ is induced by the following term in chiral perturbation (see the $L_{10}$ term in Ref.~\cite{Gasser:1984gg}),
$(g_s^2/16\pi^2 f_S^2) \phi \phi_8^a G^b_{\mu\nu} G^{c\mu\nu} d_{abc}$.
However, this three body decay receives phase space suppression compared to the decays to the SM gauge bosons such as $\phi_8 \to gg$
and can be neglected.

For a future prospect, rescaling the current sensitivity of the jet+$\gamma$ resonance search for $m_{\phi_8}=1.5$~TeV~\cite{Aad:2015ywd} to that of $\sqrt{s}$=14~TeV with 3000~fb$^{-1}$, the octet scalar with $f_S(N/3)^{-1}\lesssim500$~GeV can be proven at the LHC.

\subsection{The triplet pNG boson}

We next consider the triplet scalar $\phi_3$ briefly.
Since $\phi_3$ has the hypercharge $-4/3$, it cannot decay into the SM gauge bosons.
Then, the triplet decays into the SM fermions via the following dimension 6 operator,
$\Delta {\mathcal L}_{\phi_3} \sim (\kappa'_{ij}/\Lambda'^2) (\bar\psi  \chi) d_{R,i}^c e_{R,j}^c$,
where $\kappa'_{ij}$ are coupling constants, $\Lambda'$ is some mass scale, and 
the indices $i,j$ here denote three generations of the SM fermions.
If we take a sufficiently small $\Lambda'$, the triplet $\phi_3$ decays promptly in the LHC experiment.
The constraint on $\phi_3$ depends on the detailed flavor structure of the $\kappa'_{ij}$ couplings.
When $\phi_3$ mainly decays into an electron or a muon and a light quark,
the present upper bound on the mass $m_{\phi_3}$ is around 1 TeV \cite{Aad:2015caa}.
If $\phi_3$ mainly decays into a tau lepton and a bottom quark,
the bound becomes slightly weaker and is around 750 GeV~\cite{hKhachatryan:2014ura}.

\subsection{The baryons}

Finally, we look at the baryons in the present model.
The mass of the light baryons can be estimated by the scaling-up argument as \cite{Chivukula:1989qb}
\begin{equation}
\begin{split}
m_B &\simeq m_p \frac{f_S}{f_\pi} \sqrt{N\over 3} 
\simeq 4~{\rm TeV} \times \sqrt{N\over 3}\left(\frac{f_S}{400~{\rm GeV}}\right)  ,
\end{split}
\end{equation}
where $m_p$ is the proton mass.
The mass parameters of $M_\psi$ and $M_\chi$ generate mass splitting between the light baryons
and the lightest baryon is $\chi^N$ for $M_\psi > M_\chi$.
The decay of this baryon is induced by the following higher dimensional operator,
$\Delta {\mathcal L}_B \sim (\kappa''_{i_1\cdots i_N}/\Lambda''^{3N-4}) \bar\chi^N e^c_{R,i_1} \cdots e^c_{R,i_N}$,
where $\kappa''_{i_1\cdots i_N}$ are coupling constants, and $\Lambda''$ is some mass scale.
If we take a sufficiently small $\Lambda''$,
the lightest baryon $\chi^N$ decays before the era of the Big Bang Nucleosynthesis.
The detailed analyses are left for future investigations.

\section{Conclusion}

In this paper, from the viewpoint of the prevailing nature of new strong dynamics,
we have pursued a possibility of a new asymptotically free gauge theory to be explored at the LHC.
Motivated by the reported event excess in diphoton invariant mass distribution,
we have discussed a scenario that the lightest color-singlet pNG boson of a new QCD-like theory
is produced by gluon fusion and decays into a pair of the SM gauge bosons.
Despite the strong dynamics, the production cross section and the decay widths are calculated by the chiral anomaly.
The excess can be explained by the pNG boson with mass of around 750 GeV. The pNG boson will yield $Z\gamma$ resonance.
The model also predicts exotic hadrons such as
color octet and triplet scalars and baryons some of which are within the reach of the LHC experiment.

\begin{acknowledgments}
We thank Kenta Hotokezaka, Yevgeny Kats and Tomer Volansky for useful discussions.
YN is supported by a JSPS Fellowship for Research Abroad.
RS and KT are supported in part by JSPS Research Fellowships for Young Scientists.
\end{acknowledgments}

\bibliography{ref}
\bibliographystyle{utphys}

\end{document}